\documentclass[referee,a4paper,12pt,traditabstract, fleqn,runningheads]{swsc} 

\usepackage{graphicx}
\usepackage{txfonts}
\usepackage{subfigure}
\usepackage{epstopdf}
\usepackage[authoryear,round]{natbib}
\usepackage[backref]{hyperref}
\usepackage{url}
\hypersetup{colorlinks=vìtrue,citecolor=cyan,urlcolor=cyan,linkcolor=blue}

\begin{document}
   \title{Recurrence Quantification Analysis of Two Solar Cycle Indices}
   \author{M. Stangalini$^{1}$, I. Ermolli$^{1}$, G. Consolini$^{2}$, F. Giorgi$^{1}$ }
   \institute{$^{1}$ INAF-Osservatorio Astronomico di Roma, 00078 Monte Porzio Catone (RM), Italy\\
   $^{2}$ INAF-Istituto di Astrofisica e Planetologia Spaziali, 00133 Roma, Italy\\
   \email{marco.stangalini@inaf.it}}

  \abstract 
{Solar activity affects the whole heliosphere and near-Earth space environment. It has been reported in the literature that the mechanism responsible for the solar activity modulation behaves like a low-dimensional chaotic system. Studying these kind of physical systems and, in particular, their temporal evolution requires non-linear analysis methods. To this regard, in this work we apply the recurrence quantification analysis (RQA) to the study of two of the most commonly used solar cycle indicators; i.e. the series of the sunspots number (SSN), and the radio flux 10.7 cm, with the aim of identifying possible dynamical transitions in the system. A task which is particularly suited to the RQA. The outcome of this analysis reveals the presence of large fluctuations of two RQA measures; namely the determinism and the laminarity. In addition, large differences are also seen between the evolution of the RQA measures of the SSN and the radio flux. That suggests the presence of transitions in the dynamics underlying the solar activity. Besides it also shows and quantifies the different nature of these two solar indices.\\
Furthermore, in order to check whether our results are affected by data artifacts, we have also applied the RQA to both the recently recalibrated SSN series and the previous one, unveiling the main differences between the two data sets. The results are discussed in light of the recent literature on the subject.}
\keywords{Sun: activity}
\authorrunning{M. Stangalini et al.}
\titlerunning{Recurrence analysis of solar cycle}
\maketitle

\section{Introduction}
The impact of solar magnetism and its activity cycle on the heliosphere and near-Earth space  is nowadays well recognized. Magnetic fields generated by dynamo processes in the interior of the Sun \citep{2014ARA&A..52..251C, 2014SSRv..tmp...55K}, and emerging to the solar atmosphere, modulates the flux of particles, radiation, and magnetic field in the heliosphere. In our current technology-dependent society, the impact of heliospheric changes driven by the solar activity is becoming increasingly important \citep{baker2008solar,hathaway2004sunspot, jones2012influence}, as well as the need for accurate forecasting of the conditions in the heliosphere \citep{hapgood2012astrophysics, schrijver2015socio}.\\ 
Solar activity clearly shows a mean period of about eleven years, and variations in amplitude occurring on time scales longer than the main period. \\
Non-linear analysis methods applied to different solar indices have shown that the solar activity cycle behaves  as a low-dimensional chaotic and complex system \citep{2009A&A...506.1381C, 2010A&A...509A...5H, 2013A&A...550A...6H, zhou2014low}. Hence, linear data analysis techniques, like Fast Fourier Transform (FFT) or wavelet, applied to solar indices time series, can fail to give a complete description of the process represented by the investigated data. This is because in such techniques, non-linearities are not preserved, meaning that a fundamental property of the system (i.e. its non-linear behavior), cannot be studied at all. For a complete description and analysis of the shortcomings of applying linear techniques to non-linear systems, we refer the reader to \citet{Huang903}.\\
To this regard, the analysis of the solar cycle indices in the phase space allowed a significant step forward in the understanding of the solar activity and its underlying dynamics \citep[see for instance][]{2009A&A...506.1381C, 2013A&A...550A...6H}. Although, studying  dynamical processes and gathering physical information from their phase space embedding is generally not straightforward, robust non-linear techniques are available nowadays. In particular, over the last $30$ years, a method of non-linear data analysis has been developed to quantify the information contained in the phase space representation of a dynamical system. This method, which is called Recurrence Quantification Analysis \citep[RQA,][]{marwan2003encounters}, is based on the analysis of the recurrence plots \citep[RP,][]{eckmann1987recurrence} derived from the phase portraits of data series. Recurrence plots are diagrams representing in a 2D plot the distance between couples of states in the phase space, thus representing the recurrences of a system, a general property of dynamical systems already noticed by \citet{poincare1890probleme}.\\
    \begin{figure}
   \centering
   \subfigure{\includegraphics[width=9cm, clip]{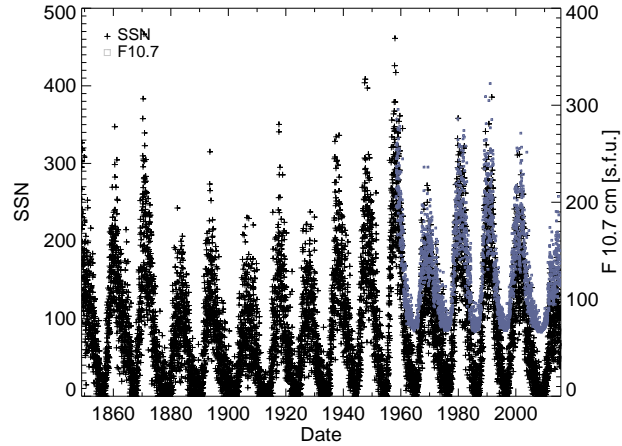}}
   \caption{Time series of the weekly averaged values of SSN2 and F10.7 analysed in our study.} 
    \label{timeseries}
   \end{figure} 
Many authors have already studied RPs of different solar indices \citep[e.g.][]{sparavigna2008recurrence, deng2015nonlinear, ghosh2015signature}, or used them to investigate periodicities and hemispheric phase relationships of the indices themselves \citep[see for instance][]{zolotova2007synchronization, li2008periodicity, zolotova2010secular, deng2013relative}.\\
RQA was also successfully employed in the analysis of non-linear systems in many different research fields \citep[see for example][for a complete review of the topic]{marwan2007recurrence} and, in particular, to uncover their dynamical transitions. However, as far as the authors know, in solar physics and heliophysics, this technique was only applied to study the temporal evolution of the deterministic states of the solar activity \citep{2002ESASP.506..197P}. In particular, the latter authors have applied the RQA to the SSN, with the aim of studying the intermittent nature of the solar activity cycle. The most interesting finding of their study was that, during the increasing phases of solar activity cycles, the determinism of the system is reduced, and this correlates with the high-frequency fluctuations of the SSN data. This is evidence for an increase of the intermittency of the analysed system, a feature which is easily uncovered by the RQA.\\
In this study, we extend the application of the RQA technique to solar data, by also investigating the laminarity of the solar cycle as represented by the most commonly used time series of solar indices, namely the SSN and the radio flux at 10.7 cm (hereafter also F10.7). Our primary goal is the study of the RQA measures in time, and highlight possible differences between the two solar indicators.  In addition, the results of this analysis may also advance our understanding of the underlying dynamics of the solar cycle, and provide useful information to be incorporated in numerical models and simulations. Our investigation can be also regarded as a timely examination of solar activity data available nowadays. Indeed, very recently, the complete sequence of the SSN has been revised to account for several calibration issues that were identified in recent years \citep{2014SSRv..186...35C, 2015arXiv151006928F}. In this work we applied the RQA on both the previous and the new SSN data series, hereafter also SSN1 and SSN2 respectively. While this is done in order to test the sensitivity of our results to the impact of data inaccuracies and artifacts, this analysis is of more general interest, serving as a non-linear comparison of the two SSN solar series.
   \begin{figure*}[!ht]
   \centering
   \subfigure[RP SSN2]{\includegraphics[width=8cm, clip, trim={0cm 3 3 3}]{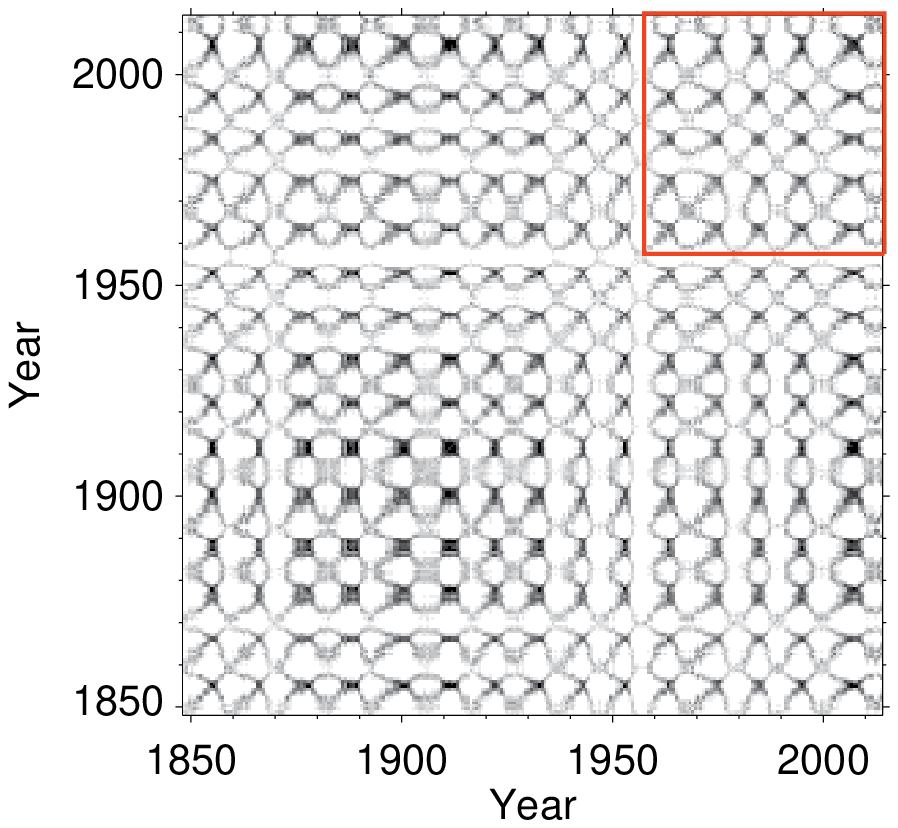}}
   \subfigure[RP F10.7]{\includegraphics[width=8cm, clip]{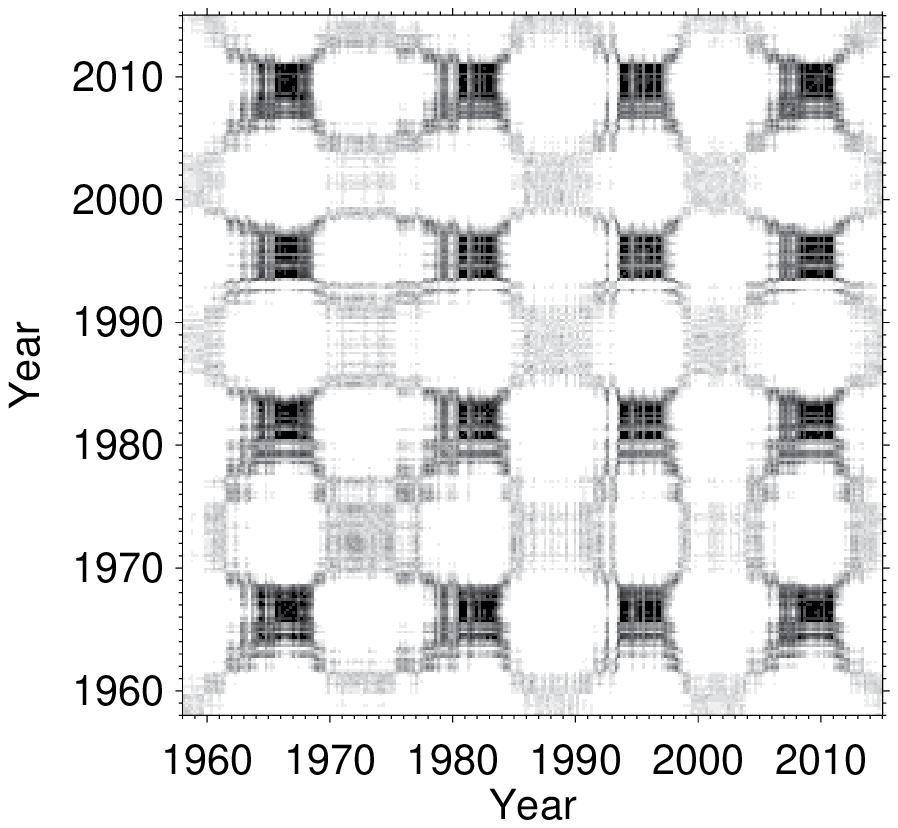}}
      \subfigure[RP SSN2 low frequency]{\includegraphics[width=8cm, clip]{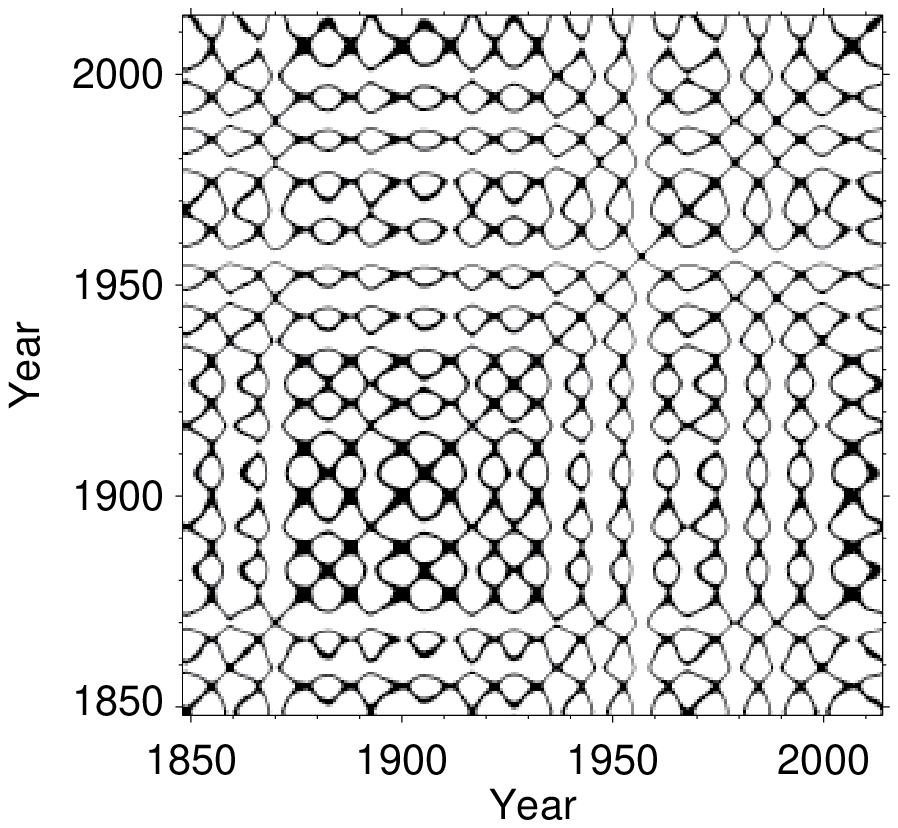}}
      \subfigure[RP SSN2 high frequency]{\includegraphics[width=8cm, clip]{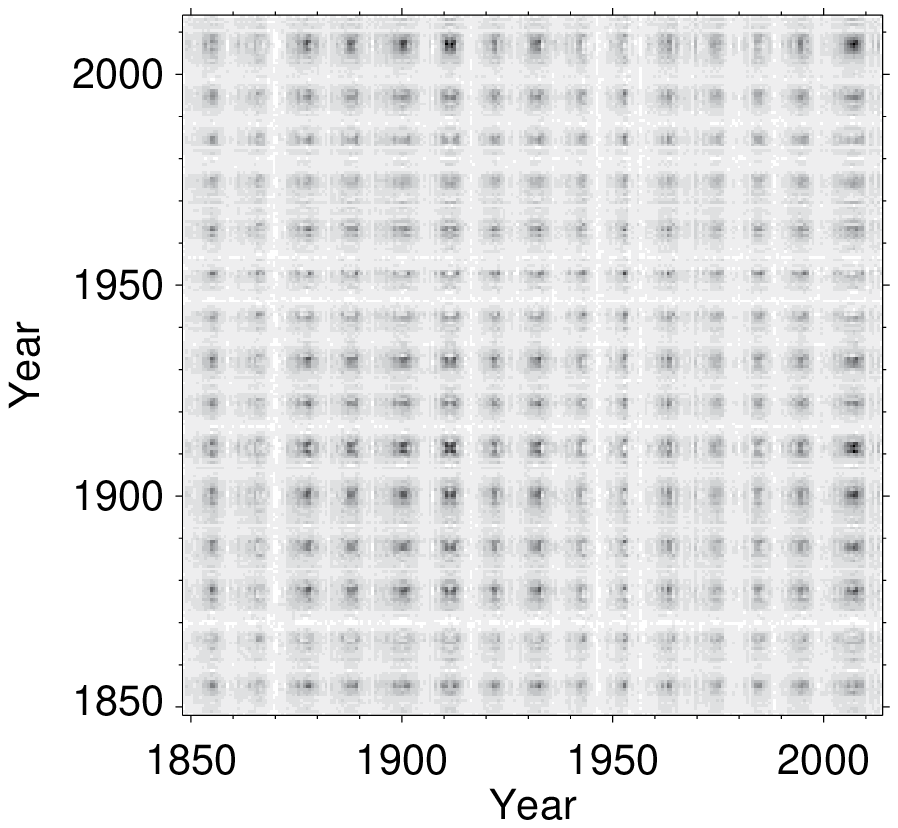}}
   \caption{(panel a): RP of the weekly unfiltered SSN2. (panel b): RP of the weekly unfiltered F10.7. The red box in the RP of the SSN identifies the period associated with the RP of the F10.7. (panel c): RP of the low-frequency part of SSN2. (panel d): RP of the high-frequency part of SSN2.}
    \label{rptot}
   \end{figure*} 

\section{Data set} 
Several indices have been introduced in order to represent the many different observables modulated by the solar cycle \citep{2010LRSP....7....1H,ermolli2014solar}. \\
The data analysed in our study consist of two time series, which are by far the most widely employed in the literature, the SSN \citep[see e.g.][]{2014arXiv1407.3231C} and the solar radio emission at 10.7 cm \citep[and references therein]{tapping201310}.\\
The SSN is defined accordingly to the formula introduced by \citet{wolf1851universal} as $SSN=k (10G + N)$ where $G$ is the number of sunspot groups, $N$ is the number of individual sunspots in all groups visible on the solar disk from visual inspection of the solar photosphere in white-light integrated radiation,  and $k$ denotes a correction factor that compensates for differences in observational techniques and instruments used by the observers in time. \\
SSN constitutes one of the longest continuous measurement programs in the history of science. It is available since 1749 and although the  series suffers discontinuities and uncertainties, it continues to be used as the most common index to describe and study solar cycle properties. As already mentioned, very recently, the SSN has been scrutinized and the series has been significantly revised \citep{cliver2013recalibrating, 2014arXiv1407.3231C, lefevre2014survey, 2015arXiv151006928F, cliver2015recalibrating} to account for discontinuities due to instrumental and observational practices. This led to the recent release (1st July 2015) of a new SSN data series. The revised sequence is available at SILSO \footnote{SILSO, http://www.sidc.be/silso/datafiles}. For detailed information on this series we refer the reader to the review of \citet{2014SSRv..186...35C}. In this work we use this new data series, as well as the previous one for comparison, to test our results against the effects of the data revision. We restrict our attention to the analysis to the longest uninterrupted SSN record spanning the last 167 years from 1849 to 2015 \citep{sidc}. This is done in order to avoid gaps that are not suited to the RQA. Indeed, the presence of gaps would inject disruptions in the RPs that might be difficult to handle. In this regard, the use of gap-filling numerical techniques \citep[see for instance][]{de2011method} may allow the extension of the RQA to earlier epochs of the sunspot record, where the analysis of dynamical transitions may offer interesting insights. However, in order to apply the RQA on such data series, a detailed analysis of the effects of gap-filling methods on the RPs and RQA measures themselves is needed.\\
It is important to remark that SSN is not a physical quantity, for this reason we complemented this data with uninterrupted weekly-averaged measurements of the solar radio flux from 1958 to 2015 \citep{benz20094, 2011SoPh..272..337T}. These measurements result from the synoptic observations of the solar radio emission made at the various observatories since 1945 \citep{sullivan2005early} at  different frequencies, ranging from 0.1 to 15 GHz. \\
Among the various radio measurements, we analysed those pertaining the flux in the wavelength range of 2.8 GHz or, equivalently, 10.7 cm, near the peak of the observed solar radio emission, made by the National Research Council (NRC) of Canada from 1947 to 1991 in Ottawa and thereafter in Penticton. These measurements constitute the longest, most stable and well-calibrated, almost uninterrupted record of direct physical data of the solar activity available to date \citep{svalgaard2010solar}. The solar radio flux is measured using the Solar Flux Unit (SFU, $10^{-22} Wm^{-2}Hz^{-1}$). A preliminary analysis of the effect of the averaging temporal window (not shown here) demonstrated that, as far as localization of the dynamical transitions is concerned, the choice of a weekly-average of the data represented a good tradeoff between the signal-to-noise ratio and the number of samples used. Although the F10.7 record is available since 1947, we noted that before 1958 a number of null samples was present. For this reason, and in order to be safe, we focused on the data starting from 1958 where the number of unavailable measurements was much more limited.\\
Figure \ref{timeseries} shows the weekly SSN2 and F10.7 values analysed in our study. Specifically the weekly averages were obtained with a seven-day average of the daily values. The SSN (both SSN1 and SSN2) and F10.7 data analysed derive from the  archives at the Sunspot Index and Long-term Solar Observation Centre at the Royal Observatory of Belgium and Canadian Space Weather Forecast Centre \footnote{CSWFC, http://www.spaceweather.ca/solarflux/sx$-$5$-$eng.php}, respectively. The series analysed in this study  were retrieved in September 2015. 

   \begin{figure*}[!ht]
   \centering
    \subfigure{\includegraphics[width=8cm, clip]{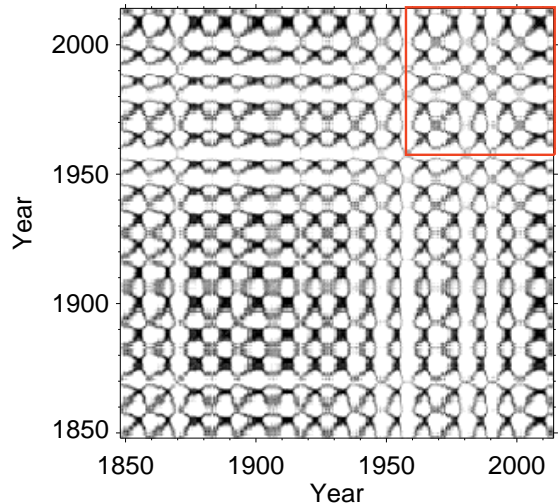}}
   \caption{RP of of the weekly unfiltered SSN1 (no embedding).}
    \label{RPv1}
   \end{figure*} 

\section{Methods}
\begin{figure}[h]
 \centering
   \subfigure{\includegraphics[width=7cm, clip]{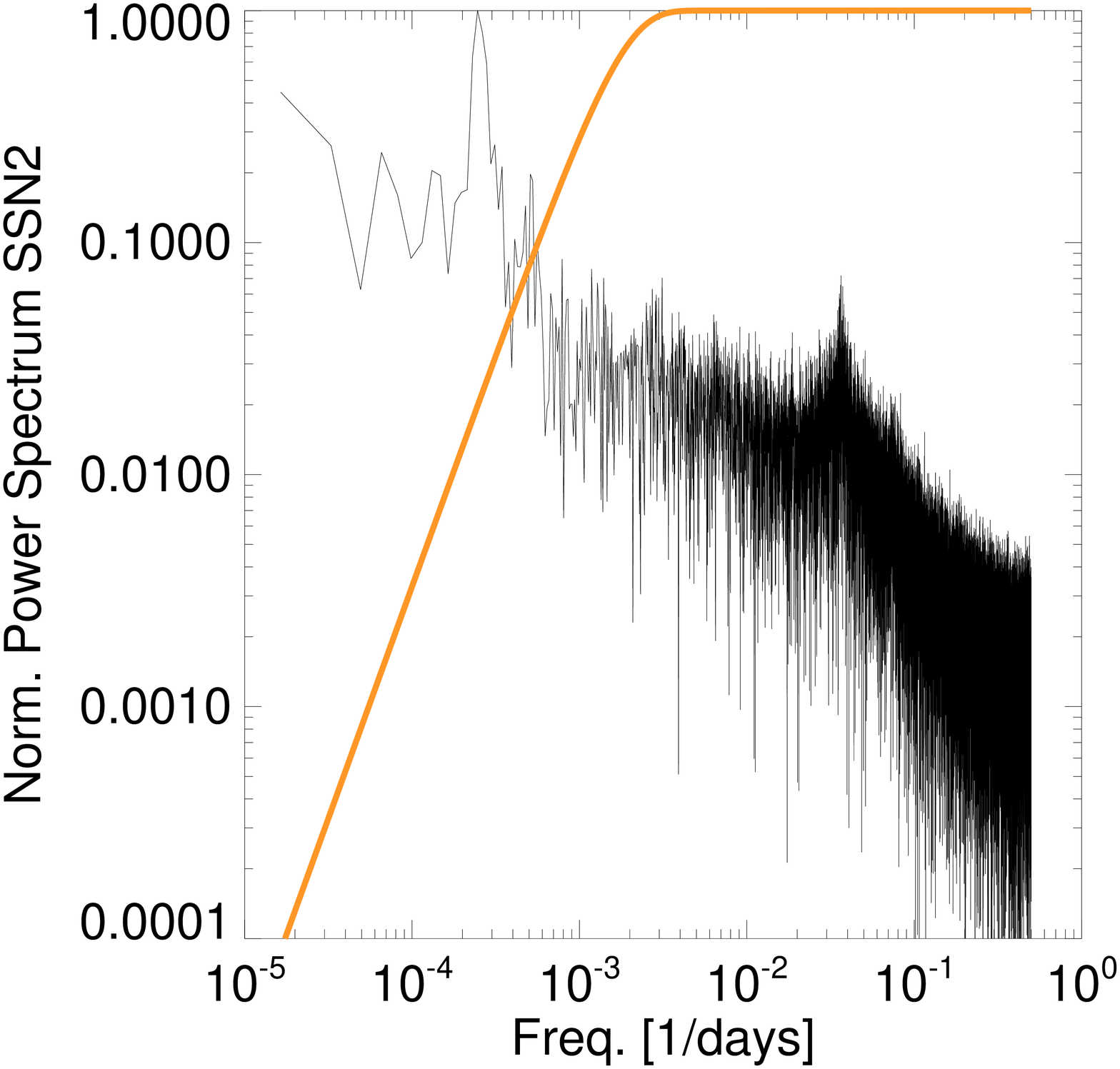}}
 \subfigure{\includegraphics[width=8cm, clip]{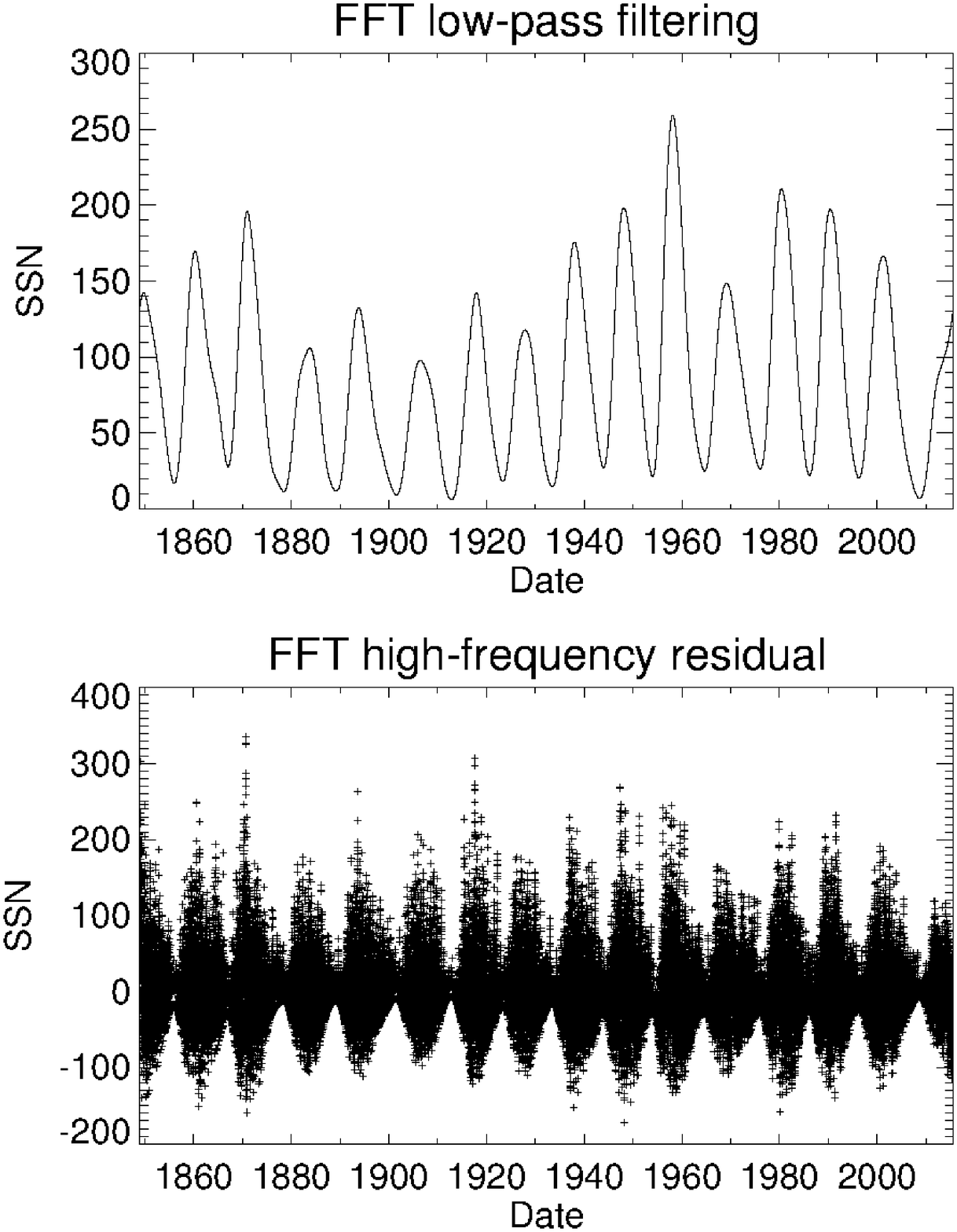}}
 \caption{Power spectrum of SSN2 and high-pass filter (left panel). Reconstructed low-frequency (upper right) and high-frequency (lower right) part of SSN2.} 
 \label{FFTfiltering}
 \end{figure}

\begin{figure}[h]
 \centering
 \subfigure{\includegraphics[width=11cm, clip]{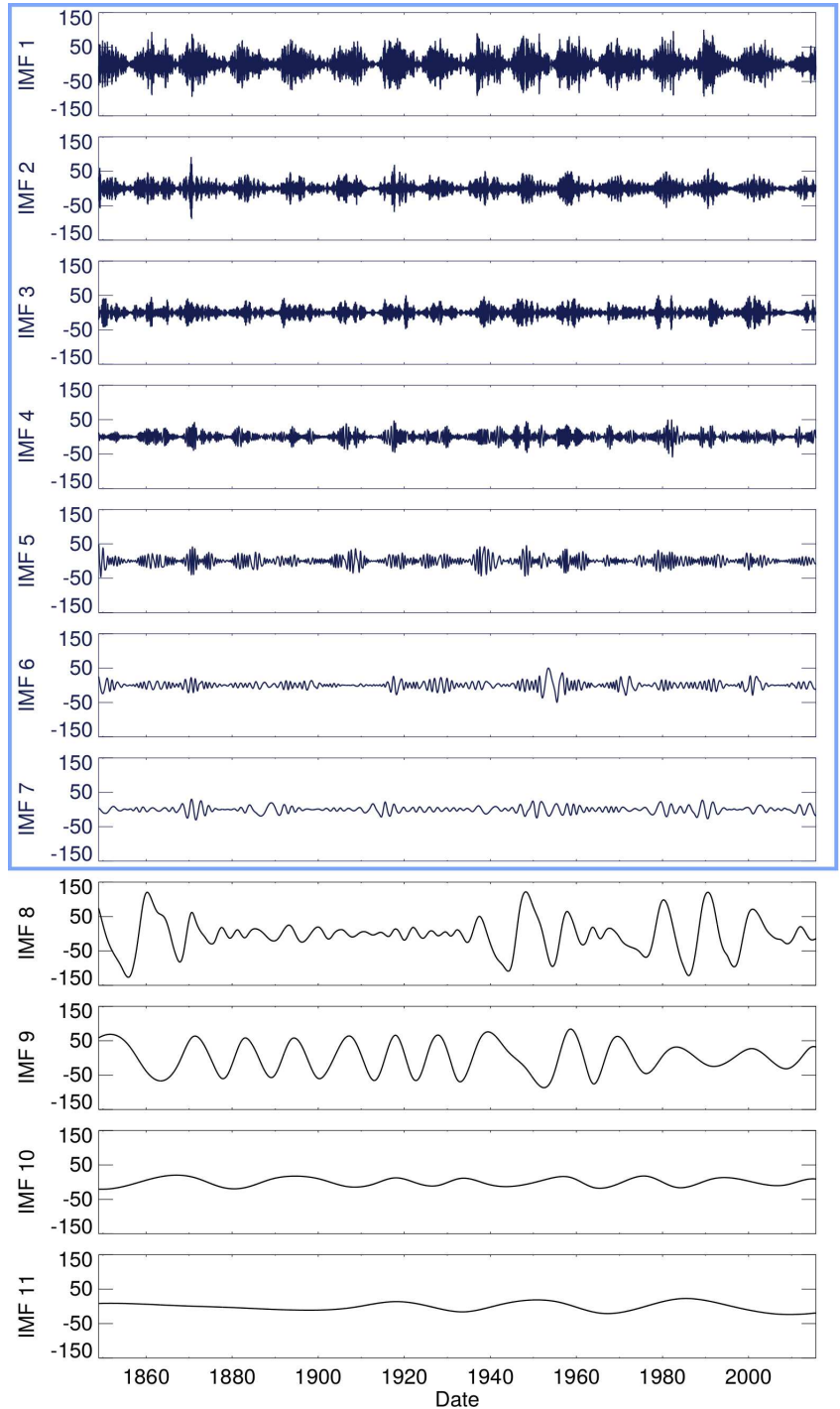}}
 \caption{EMD decomposition of the SSN2 data sequence. The first seven IMFs (blue box) are used as a representation of the high-frequency part of the signal.} 
 \label{IMFs}
 \end{figure}

\begin{figure}[h]
 \centering
 \subfigure{\includegraphics[width=9cm, clip]{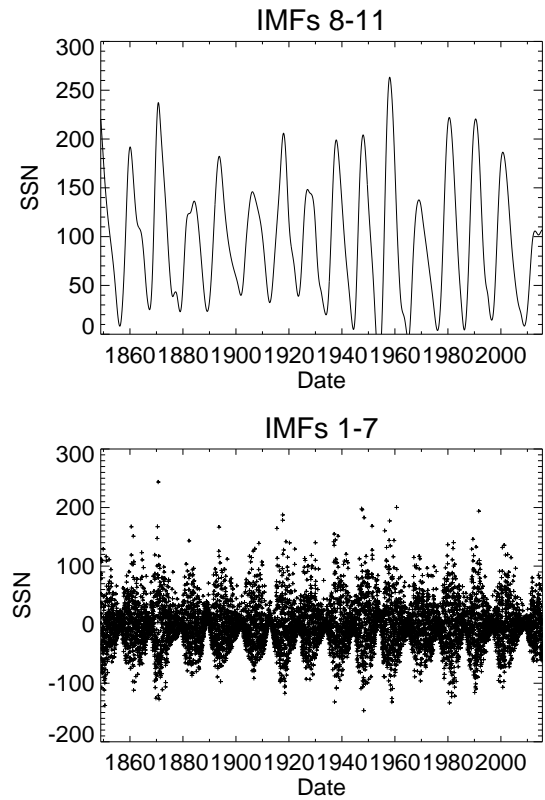}}
 \caption{EMD reconstruction of the low- (top) and high-frequency (bottom) part of the SSN2. The high-frequency part of the signal is computed by co-adding the first seven IMFs, while the low-frequency part, not used in the RQA, by adding the remaining ones.}
 \label{EMDfiltering}
 \end{figure}

The starting point of our analysis is the evaluation and study of the RPs from the phase reconstruction of the SSN2 and F10.7 series. Introduced by  \citet{eckmann1987recurrence}, RPs are diagrams  that visualize the trajectory of the system, represented by the analysed data series, in a 2D domain \citep{iwanski1998recurrence}. In RPs, each point (i,j) of the diagram is shaded according to the distance between two points $X_{i}$ and $X_{j}$ on the trajectory in the phase space. The closeness of the states of the system at different times (recurrences) determines specific features and cluster of points in the plot, which describe the nature of the dynamical system. Indeed, recurrences are a characterizing property of any dynamical system \citep{kac1947notion}. \\ 
Figure \ref{rptot} shows the thresholded RPs derived from the SSN2 (upper left) and F10.7 (upper right). For the seek of simplicity, the embedding parameter is $m=1$ (no embedding), while the time delay parameter used to produce the phase portrait, from which the RP are obtained, is $\tau=1$. Indeed, \citet{iwanski1998recurrence} has shown that, qualitatively, features of RPs generated from high embedding dimensions, are also seen when using small embedding dimensions. This point will be further discussed in the next sections. The threshold used to construct the RPs is $15$. In Fig. \ref{RPv1} we show for comparison the RP of the unfiltered SSN1.
The convention adopted throughout the manuscript is such that a recurrence state is marked by a black dot in the RPs. It is worth mentioning that transitions markers in the RQA measures are rather insensitive to the exact choice of the embedding parameters, as demonstrated by \citet{iwanski1998recurrence}. The main features of plots of Fig. 2 are discussed in section 4. \\
In order to extract quantitative information from RPs, we applied the RQA technique \citep[see for example][]{zbilut1992embeddings,trulla1996recurrence, thiel2004much, webber2005recurrence}. This method is based on the analysis of the distributions of recurrence points in the vertical lines and the diagonal lines of RPs \citep[for a review see e.g.][]{marwan2003encounters, marwan2007recurrence}. Indeed, diagonal lines in RPs identify trajectories that regularly visit the same region of the phase space at different times. This is a characteristic feature of deterministic systems. For this reason, the lenght of the diagonal lines in RPs represents a measure of determinism (DET). In contrast to this, vertical lines in RPs mark states which are trapped for some time. Thus the lenght of vertical lines in RPs can be regarded as a measure of the laminarity of the system (LAM). \\
The RQA also allows the study of other complexity indicators of dynamical systems, although these are not relevant to our aim \citep[for a review see e.g.][]{marwan2007recurrence,zbilut2006recurrence}. In this work, we restricted our attention to DET, and LAM. These two measures have been successfully used several times to identify dynamical transitions \citep[e.g.][]{marwan2013recurrence}.\\
In order to estimate DET and LAM from RPs, in this work we used the well tested command-line recurrence plots code, which is part of the TOCSY (Toolboxes for Complex Systems) toolbox\footnote{The command-line recurrence plots code is freely available at the following link: http://tocsy.pik-potsdam.de/} and, more in particular, its RQA utility and the time-delay embedding of the time series. For more information about the algorithms and methods employed in the toolbox (time-delay embedding, generation of RPs, and RQA application) we refer the reader to \cite{marwan2007recurrence}.
In this code the embedding is performed through a time-delay technique. Given a time-discrete measurement of an observable $u_{i}=u(i \Delta t)$, where $i=1,...,N$ and $\Delta t$ is the sampling time, the phase space can be reconstructed as follows:
\begin{equation}
\textbf{x}_{i}= \sum_{j=i}^{m} u_{i+(j+1)\tau} \textbf{e}_{j};
\end{equation}
where $m$ is the embedding dimension, $\tau$ the time delay, and $\textbf{e}_{i}$ are unit vectors spanning an orthogonal coordinate system.\\
We applied the command-line recurrence plots code to the aforementioned solar indices with a moving window of $100$ weeks and a step of $1$ week. This was done to study the evolution of the RQA measures in time.\\
In order to quantify the effects of the new recalibration of SSN2, we also made use of Joint Recurrence Plots \citep[JRPs][]{marwan2004cross} that are suited for studying the similarities between two data series in the phase space, and specifically identifying times at which they share the same recurrences. A JRP is a plot showing all the times at which a recurrence in one dynamical system occurs simultaneously with a recurrence in another dynamical system. Indeed, a JRP is the Hadamard product of two RPs representing two dynamical systems.\\ 
In Fig. \ref{rptot} (upper panels) it is clear that RPs of the time series of such two indices are almost completely dominated by the low-frequency modulation of the solar activity cycle ($11$ years). In order not to make the RQA measures biased by this modulation, both the SSN2 and F10.7 were high-pass filtered. This was done, at first, by FFT filtering the data with a filter whose cut-off was set at $2 \times 10^{-3}$ days$^{-1}$ (see Fig. \ref{FFTfiltering}). In panel (c) and (d) of Fig. \ref{rptot} we show the RP of the low-frequency and high-frequency part, respectively. However, we note that, in contrast to the RQA, the FFT technique is a linear method, thus it might not represent a suitable preconditioning technique to be used on the data. For this reason, and in order to independently check the reliability of the results, we also used the empirical mode decomposition \citep[EMD][]{Huang903} to filter out the low-frequency dynamics. This was done only on the SSN2 to check the consistency of the results obtained from the FFT filtering. Indeed, the EMD technique preserves the non-linearities of the signal, thus represents a more safe option for the pre-processing of the data examined in this work. The EMD analysis was already applied to decompose the solar cycle (to the SSN) in a series of intrinsic mode functions \citep[IMFs;][]{0004-637X-830-2-140}. It consists of an iterative process that, starting from the envelopes of maxima and minima estimates each IMF as the mean value of these two envelopes.  The signal is therefore decomposed in a sequence of IMFs which are locally defined from the signal, without making use of any predefined decomposition basis or assumption, and that offer a data-driven decomposition of the signal. 
In Fig. \ref{IMFs} we show the EMD decomposition of the SSN2 data series. By co-adding the low frequency IMFs and the high-frequency IMFs, one can decouple the eleven-year long-term periodicity from the rest. In Fig. \ref{EMDfiltering} we show the result of that, where the high-frequency part of the signal (the one used in this work) is obtained by co-adding the first $7$ IMFs (see blue box in Fig. \ref{IMFs}). The high-frequency part of the data sequences are then studied with the RQA analysis.\\
As pointed out by many authors \citep[see for instance][]{schinkel2008selection}, the choice of the optimal embedding may have some impact on the exact value of RQA measures, but only a negligible effect on the position of the markers of dynamical transitions \citep{iwanski1998recurrence}. Indeed, \cite{iwanski1998recurrence}, by analysing the RPs of well understood physical systems, have shown a good structural stability of RPs for different values of the embedding parameter. This means that, qualitatively, features in RPs are rather independent of the exact choice of the embedding. This is the case at least for those recurrence points that do not vanish in the RPs due to the change of the embedding itself. This property of RQA is recognized by the same authors as "counterintuitive". In fact, since the embedding process is employed to unfold the dynamics, one would expect a dramatic change of the RP for different embeddings, but this is not the case. The same authors have also shown that, while the position of the dynamical markers in the RPs is insensitive to the choice of the embedding, a gradual fading of the main features of RPs is observed as the embedding dimension increases. \\ 
Formally, the optimal minimum embedding dimension is linked to the dimension of the chaotic system $d$, so that $m \geq 2d+1$ \citep[see for instance][and references therein]{ma2006selection}. This implies that the dynamical system should be perfectly known before performing a phase space reconstruction. But this obviously makes a contradiction. In this regard, several methods were proposed to estimate the optimal embedding dimension $m$ \citep[see for example][]{fredkin1995method, rhodes1997false}. In particular, \citet{sello2001solar}, using the false neighbors method \citep[see for instance][]{kennel1992determining, abarbanel1993analysis}, found that the minimum embedding dimension for the sunspot number is $m=5$. This result indicates that the sunspot sequence is consistent to a low-dimensional system, in agreement with other independent works on the subject \citep[e.g.][]{1996A&A...310..646Z}. 

\section{Results}
\subsection{Analysis of the RQA measures DET and LAM}
Fig. \ref{rptot} shows the RPs derived from the SSN2 and F10.7 with no embedding ($m=1$ and $\tau=1$). These RPs display the density of the recurrence states of the dynamical system represented by the two analysed series as a function of time. The RPs show a varying density of recurrences and, although most of the dynamics appears to be deterministic (see the large presence of diagonally aligned features), sudden interruptions can also be seen, as for example between 1955 and 1960. It is worth noting here that this period was identified as an unusual solar cycle \citep{1990SoPh..125..143W, 2006A&A...447..735T}. The RP of the SSN2 also shows an interval affected by an increase of the density of recurrence points, which is located roughly between 1875 and 1940. While over long timescales, an almost constant eleven-year periodicity is evident as a repeating pattern of diagonal states, some modulation of this period can be found, for example, around 1900 as a small distortion of the diagonal features. All these elements are even more evident in the RP obtained from the low-frequency part of SSN2 (panel c), and reflect the non-stationary nature of the process represented by the analysed series. Another interesting aspect of the RPs, is the overdensity of recurrence points between $\sim 2007$ and $2010$.\\
 \begin{figure}[h]
 \centering
 \subfigure{\includegraphics[width=8cm, clip]{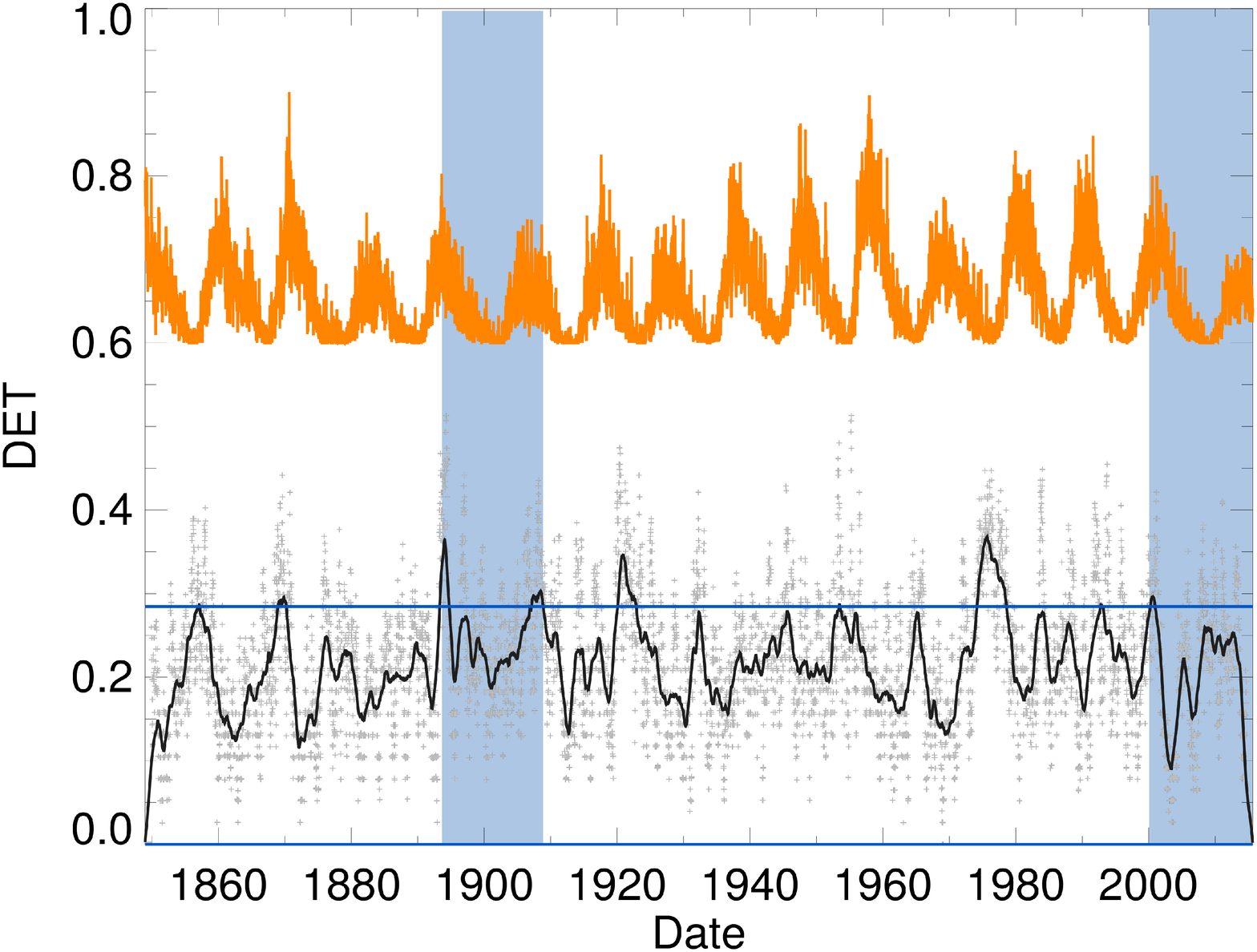}}
 \subfigure{\includegraphics[width=8cm, clip]{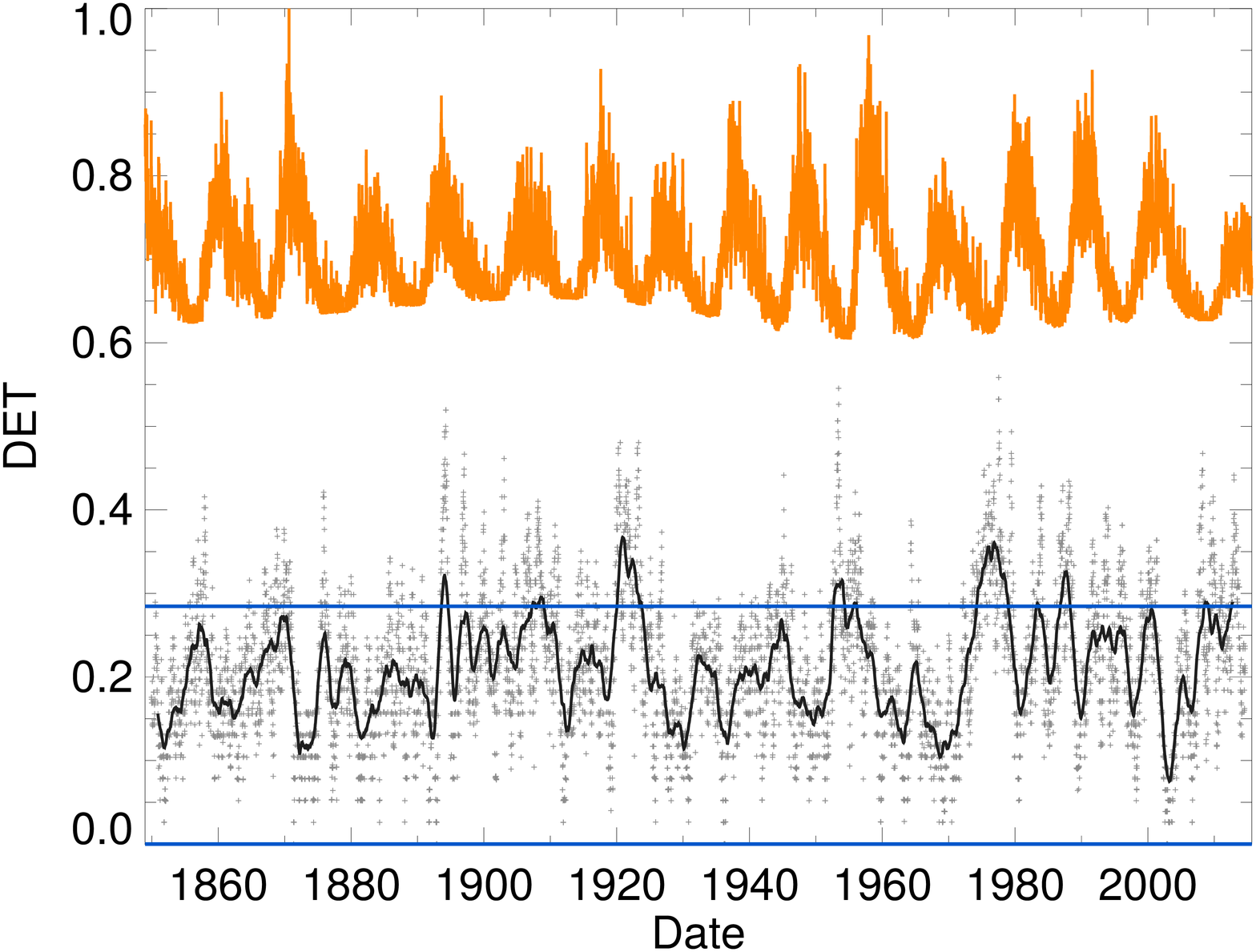}}\\
 \subfigure{\includegraphics[width=8cm, clip]{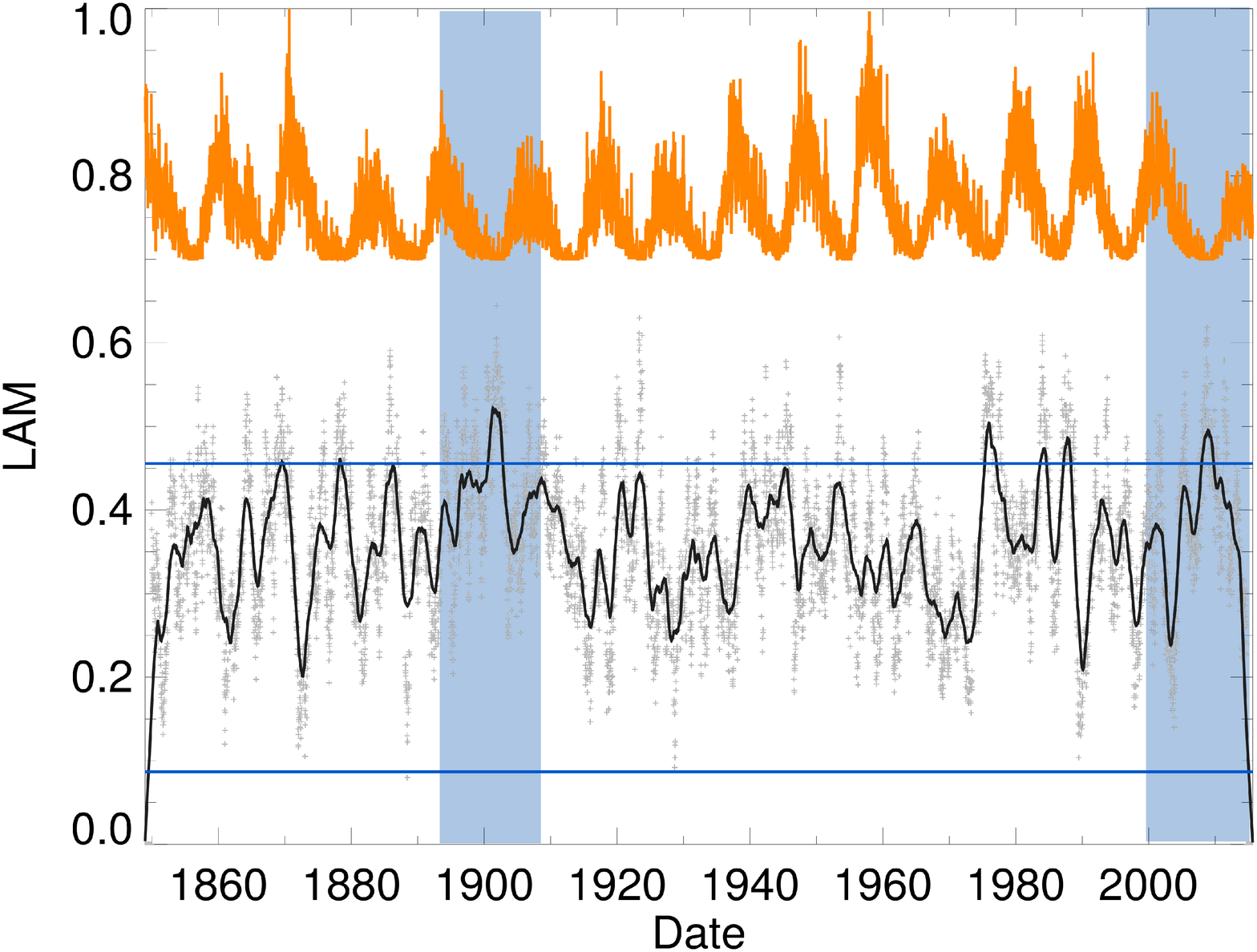}}
 \subfigure{\includegraphics[width=8cm, clip]{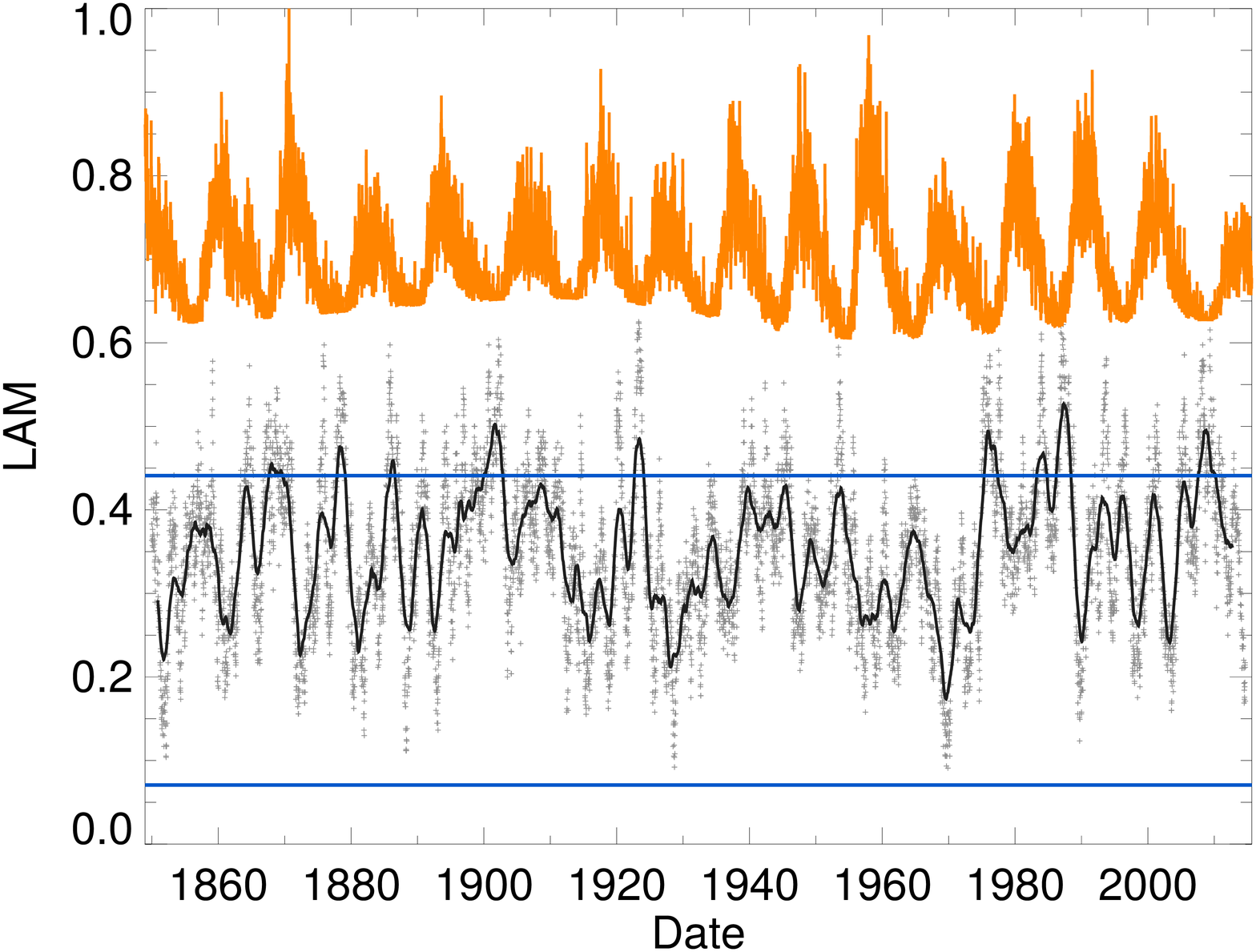}}
 \caption{Evolution of DET (top) and LAM (bottom) obtained from the SSN2 for both filtering strategies (grey dots): FFT (left panels), and EMD (right panels). The continuous line represents a smoothed version of both DET and LAM. The shaded areas highlight the times at which significant mismatches in the evolution of  DET and LAM are observed. The orange curve represents a rescaled version of the SSN2 to help the reader identifying the solar cycles. The horizontal blue lines represent the $95 \%$ confidence level.}
 \label{RQA}
 \end{figure}  
Fig. \ref{RQA} shows that the evolution of DET (upper panel) and LAM (lower panel) obtained by applying the RQA on the high-pass filtered SSN2 time series, for both the FFT (left panels) and EMD (right panels) filtering. The embedding parameters are $m=5$ and $\tau=15$. In the same plots we also display a not-to-scale version of the SSN2 (orange curve) to help the reader in the comparison of the evolution of LAM and DET with the solar activity cycle. Following \citet{marwan2013recurrence}, we used an adaptive threshold in the sliding window employed to estimate the temporal variation of the RQA measures. This is done to maintain an optimal constant recurrence rate of the order of a few percent ($10 \%$ in our case), and to keep the statistical sample constant within the sliding window. The adaptive threshold guarantees the stability of the recurrences over the temporal window considered. This is needed in order for the RQA measures not to reflect the intrinsic variations of the density of the recurrent points. Indeed, in the presence of a modulating recurrence rate, this modulation can enter the other RQA measures, preventing the analysis of the true fluctuations of DET and LAM. Since the RQA measures are statistical values estimated from the distribution of points in the moving temporal window, it is of paramount importance to dynamically adapt the threshold in such a way that the density of recurrent points in different temporal windows in the RP remains constant. In our case, the standard deviation of the recurrence rate is reduced at $0.7 \%$ by the adaptive threshold.\\
The $10 \%$ is chosen in such a way that the RQA measures do not show saturation or clipping. However, \cite{marwan2011avoid} has shown that the selection of the threshold is not critical.\\ 
As already mentioned, the size of the sliding window used in the RQA ($\sim 2$ years) is chosen to accurately sample the underlying periodicity of the solar cycle, while maintaining the computational load at a reasonable level. \\ The results obtained by using the two filtering techniques (i.e. FFT and EMD) are rather similar, and both show large fluctuations of DET and LAM. Some of these fluctuations appear in correspondence of the minima of SSN2 (see for example the peak within 1970-1980), however this is not always the case. More important, there exist specific times where LAM and DET show a different behaviour. This is the case, for instance, around $\simeq 1900$ and during the last minimum ($2005-2008$). In these two periods, in fact, LAM appears more pronounced than DET, suggesting a possible increase of the laminarity of the system. \\
In the same plots, we also show with horizontal continuous lines the $95 \%$ confidence levels as obtained from a significance test as in \citet{marwan2013recurrence}.  More in depth, the significance test is based upon a randomization (random permutation) of the SSN2 to get rid of any temporal correlation in the signal. After the randomization of the time series, the RQA is applied in order to estimate the $95 \%$ confidence level of each RQA measure. Since the resulting RQA measures obtained from the randomized signal show a non-gaussian distribution, we used the cumulative distribution function (CDF) to select the threshold corresponding to the $95 \%$ confidence level. This analysis reveals that most of the peaks of DET and LAM are statistically significant, as they exceed the upper confidence level. In contrast, none of the "negative fluctuations" (minima) exceed the lower confidence level, and can be considered statistically not significant.\\
In Fig. \ref{SSNF107compare} we compared the DET and LAM of the SSN2, with those of the F10.7 data sequence, in the time window where both measures are available (since 1958).
It is interesting to note here that, while the filtering technique has a little effect on the RQA measures, the DET and LAM of the F10.7 present large differences with respect to the SSN2 sequence. This can be noted, for example, during the last minima where the F10.7 shows a sharper peak of both LAM and DET with respect to the SSN2. In addition, it is interesting to note that most of the pronounced peaks of DET and LAM obtained from F10.7 are located in proximity of the solar minima, although some is not (see for example the peak around 1980).\\
 \begin{figure}[h]
 \centering
 \subfigure{\includegraphics[width=8.cm, clip]{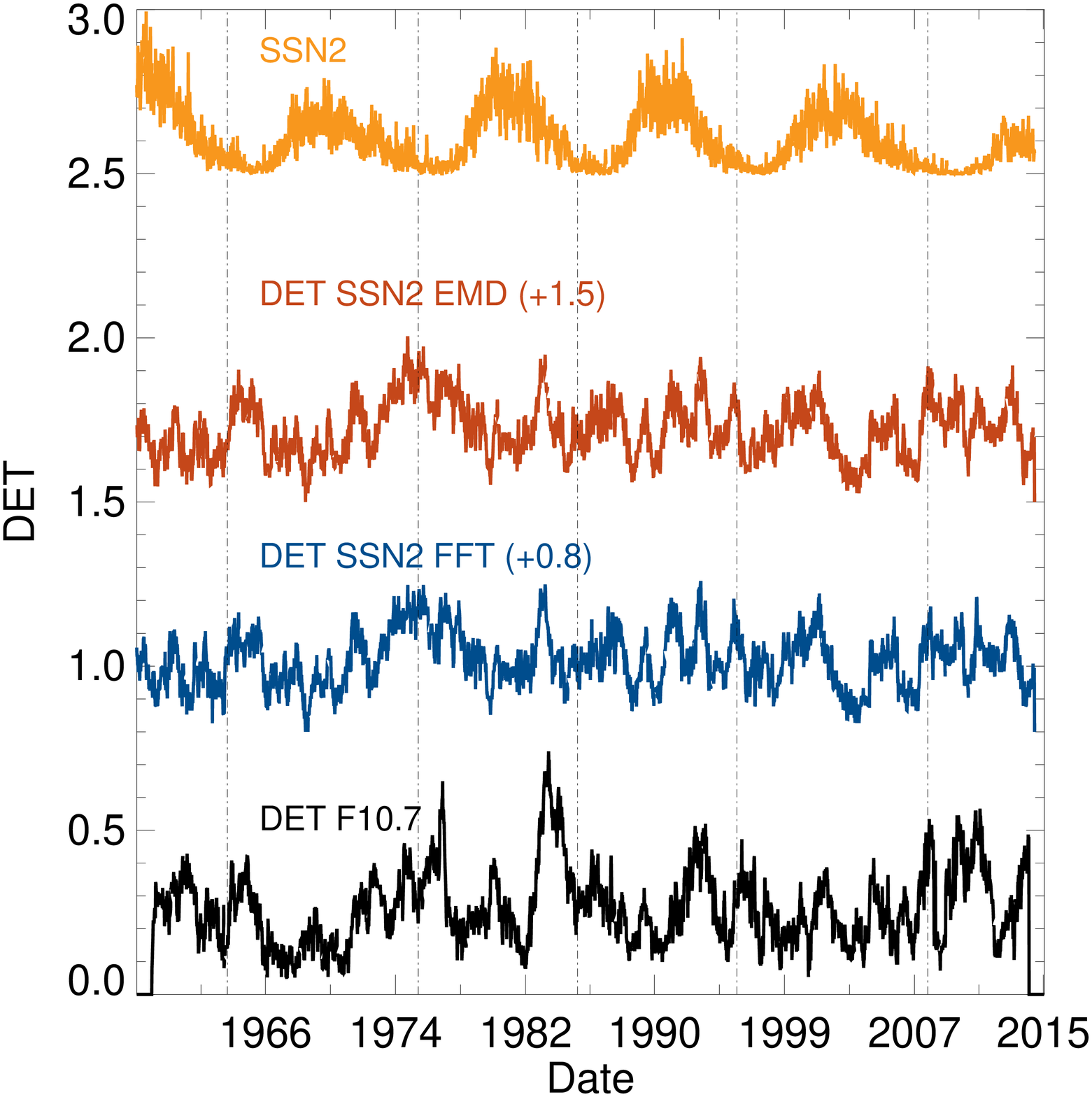}}
 \subfigure{\includegraphics[width=8.cm, clip]{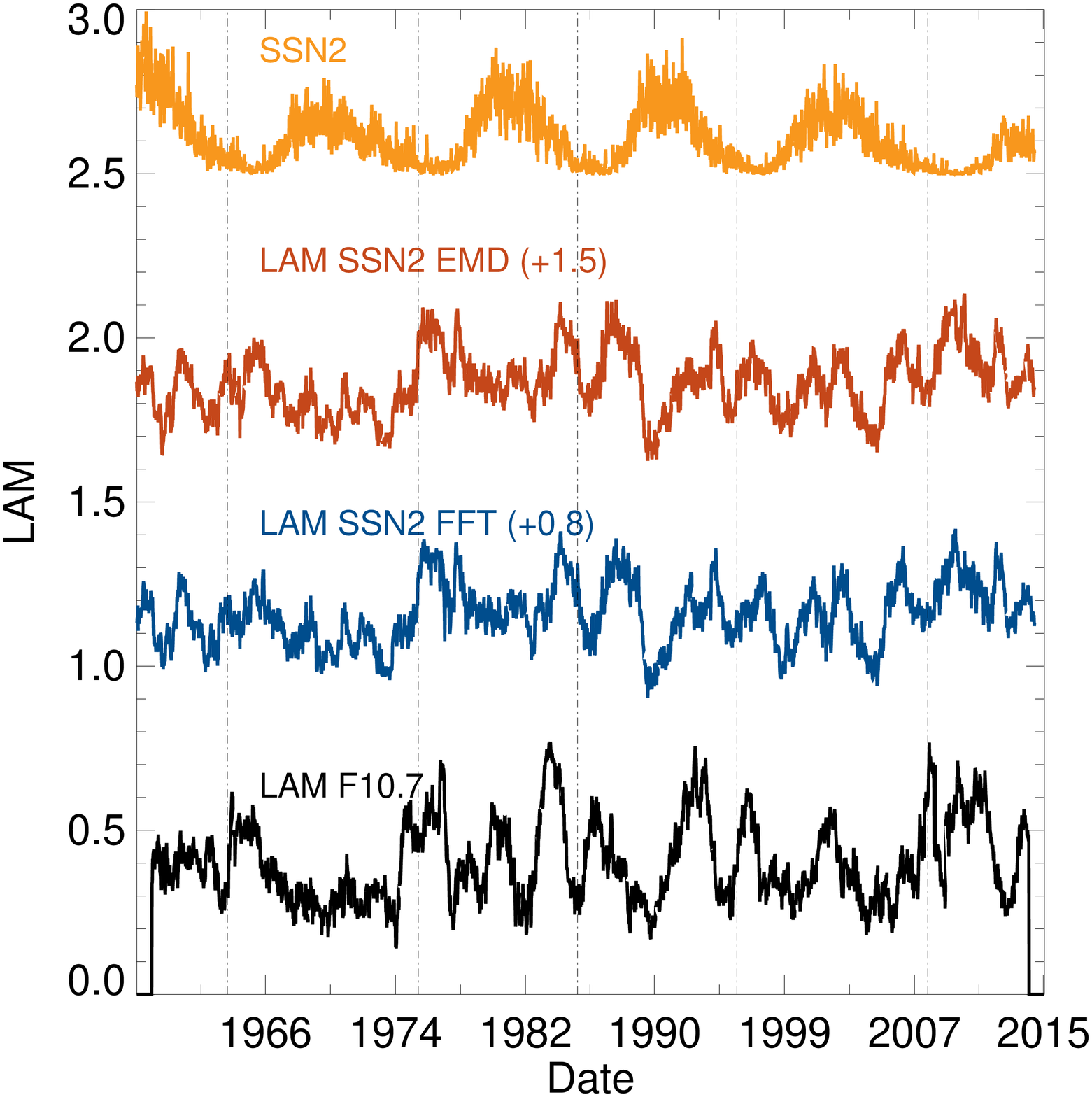}}
 \caption{\textit{Left panel:} Comparison of DET obtained from SSN2 (both filtering strategies) and F10.7 (FFT filtering). \textit{Right panel:} Comparison of LAM obtained from SSN2 (both filtering strategies) and F10.7 (FFT filtering). The dot-dashed vertical lines mark the position of the solar minima. The different data sequences have been shifted by a constant value for graphical reasons.}
 \label{SSNF107compare}
 \end{figure}

\subsection{Analysis of the fluctuations of the RQA measures}
\begin{figure}[h]
 \centering
 \subfigure{\includegraphics[width=15cm, clip]{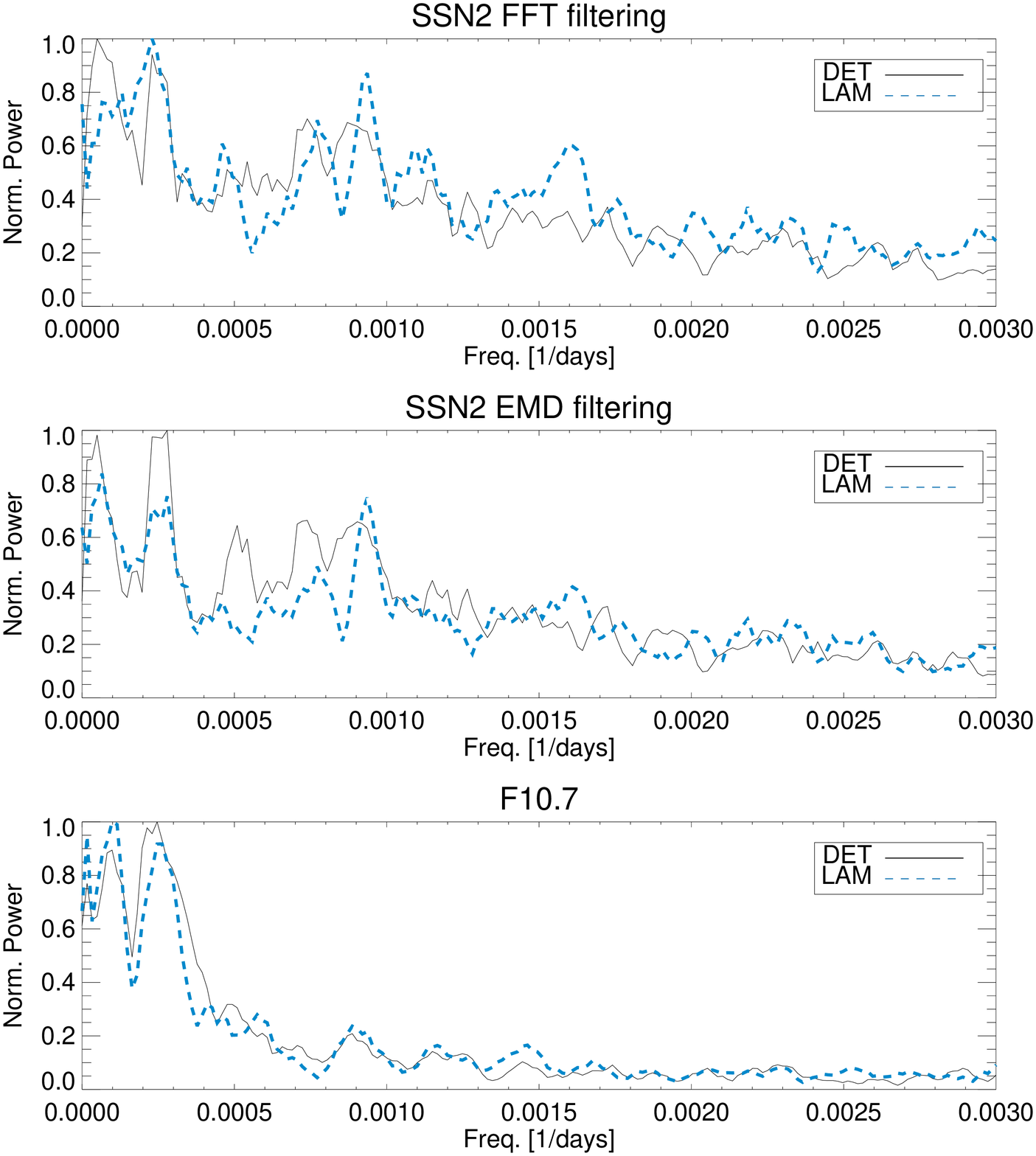}}
 \caption{Power spectrum of the DET obtained from SSN2 with FFT filtering (upper panel), from SSN2 with EMD filtering (middle panel), and from F10.7 with FFT filtering (bottom panel).}
 \label{power}
 \end{figure}

In order to investigate the oscillations seen in the RQA measures, here we study the power spectra of DET and LAM obtained from SSN2 (both filtering techniques) and F10.7 (see Fig. \ref{power}). Interestingly, the power spectra highlight differences between SSN2 and F10.7, and between DET and LAM. Indeed, while the power spectrum of DET parameter from the SSN2 shows three different peaks around $1-3 \times 10^{-4} ~days^{-1}$ (eleven-year period), $8-10 \times 10^{-4} ~days^{-1}$ (three-year period), and $16 \times 10^{-4} ~days^{-1}$ (two-year period), the last two peaks are absent in the power spectra of both LAM and DET from F10.7. In addition, restricting our attention to the power spectrum of the RQA measures from the SSN2, we note that the last peak aforementioned is much less pronounced in the spectrum of DET with respect to that of LAM. This suggest a different dynamical behaviour of the RQA measures and, more important, of the investigated solar indices. These differences will be further investigated in a future work.\\

\subsection{Comparison between the new SSN data series and the previous one}
\begin{figure}[h]
 \centering
  \subfigure{\includegraphics[width=9cm, clip]{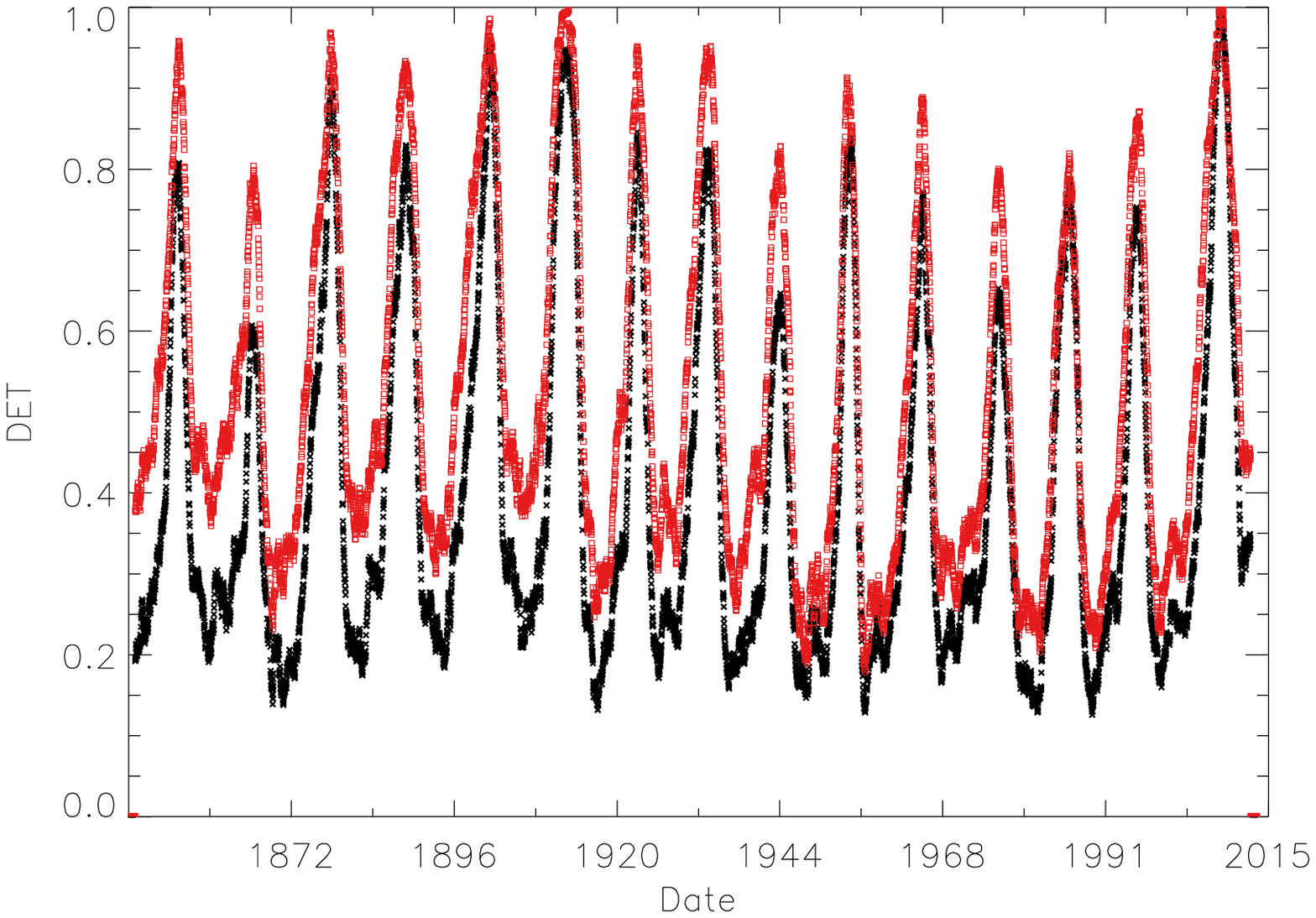}}
    \subfigure{\includegraphics[width=9cm, clip]{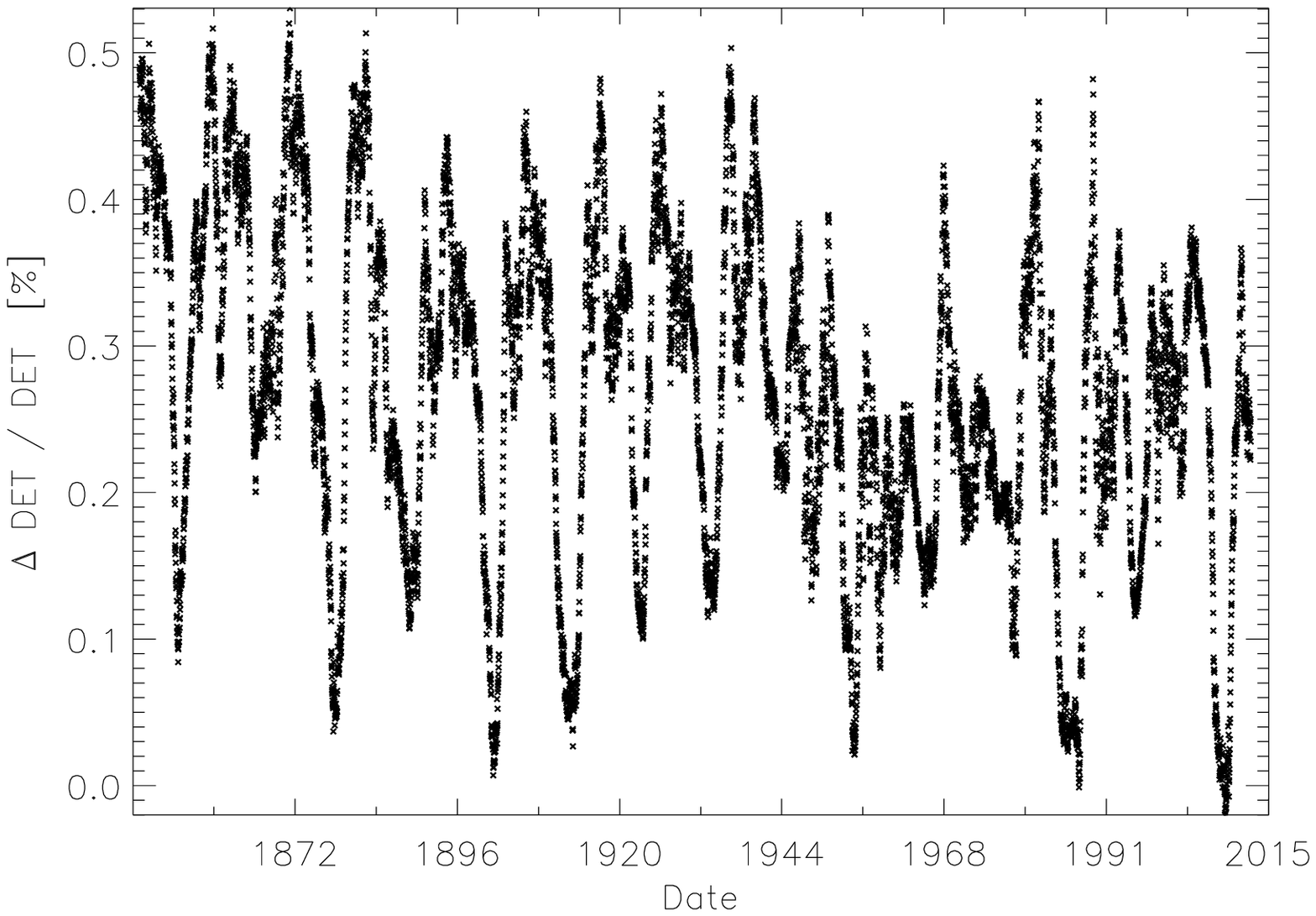}}
 \caption{Upper panel: Evolution of DET from SSN1 (black), and DET from SSN2  (red). Lower panel: relative increase of the determinism of SSN2 with respect to SSN1.} 
 \label{confronto_DET}
 \end{figure}

\begin{figure}[h]
 \centering
  \subfigure{\includegraphics[width=12cm, clip]{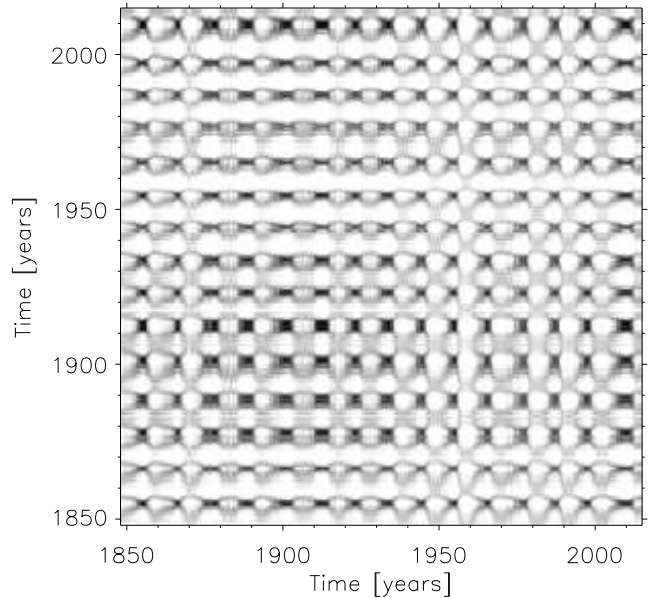}}
 \caption{JRP between SSN1  and SSN2 data series (no embedding).}
 \label{CRP}
 \end{figure}

Most of the works in the literature on large scale solar magnetism derive from the analysis, with different techniques, of the previous SSN1 series. In order to point out differences in the SSN1 and SSN2 that may be ascribed to the different calibration, we applied the RQA on both SSN1 and SSN2 series. Since we are interested in comparing the two data series, in this analysis we make use of unfiltered data. This is done in order to keep the recurrences of the system unchanged and allow the comparison of the two time series. In Fig. \ref{confronto_DET} (upper panel), we show the evolution of DET obtained from the RQA of both SSN1  and SSN2 with no embedding (i.e. embedding parameters  $m=1$ and $\tau=1$), and constant threshold $\epsilon=5$. The determinism of the SSN2 appears larger than that of SSN1. This is the case in correspondence of both minima and maxima of the solar cycle. In the same figure (lower panel) we also show the relative variation of DET of SSN2 with respect to DET of SSN1. The increase of DET is larger at the turning points of the solar cycle. In other words, the SSN2 data series shows a level of determinism significantly larger than the determinism of the previous SSN1  series; this is found at all times, but especially during the maxima of the solar cycle. In addition, this plot also shows a period of decreased DET variability between $\sim1947$ and 1980 (see Fig. \ref{confronto_DET} lower panel). We note that this period corresponds with the well-known "Waldmeier" jump, an issue that was previously identified through a comparison with the Sunspot Group Number, and finally corrected in the SSN2 release \citep{2015arXiv151006928F}.\\
However, we note that in general the recalibration of the SSN series has the only effect of changing the level of determinism of the system represented by the data, keeping the transition markers unchanged.\\
In order to better visualize the differences between the two SSN series, we made use of a JRP.\\
In Fig. \ref{CRP} we show the JRP of SSN2 and SSN1 without embedding, as in Fig. \ref{rptot}. A first look at the JRP (Fig. \ref{CRP}) and the RPs in Fig. \ref{rptot} does reveal a remarkable similarity between SSN1 and SSN2.\\ In the following we compare the two time series SSN1 and SSN2 with the F10.7 time series by using JRPs. Indeed, the comparison of the JRP obtained from SSN1 and F10.7, and the JRP obtained from SSN2 with F10.7, can reveal the presence of discrepancies between the two SSN time series much better than what the single JRP between SSN1 and SSN2 can.

\subsection{Comparison between the two SSN data series, and F10.7}
\begin{figure}[h]
 \centering
  \subfigure{\includegraphics[width=9cm,trim={0cm 0 1 12}, clip]{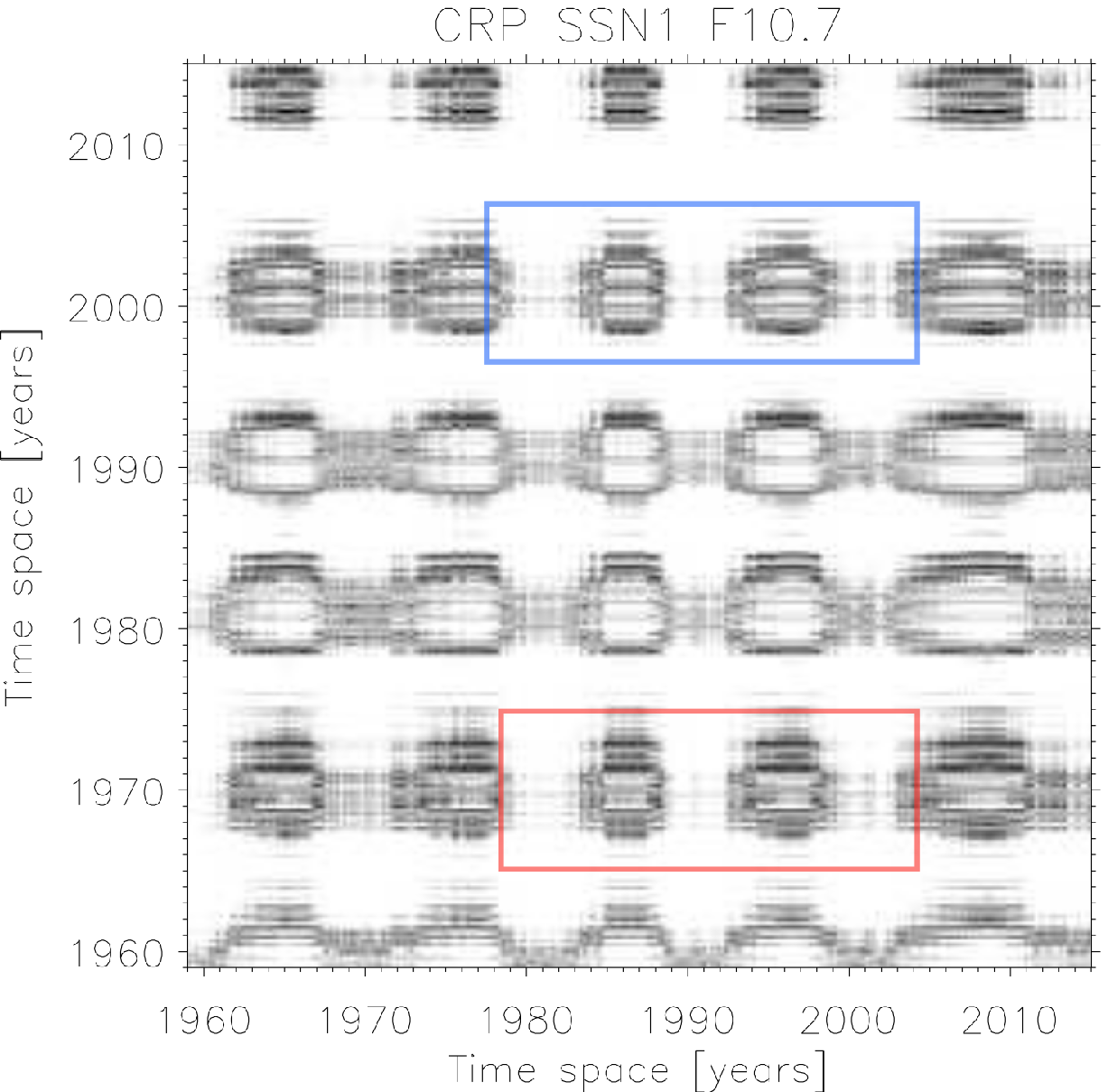}}
  \subfigure{\includegraphics[width=9cm,trim={0cm 0 1 12}, clip]{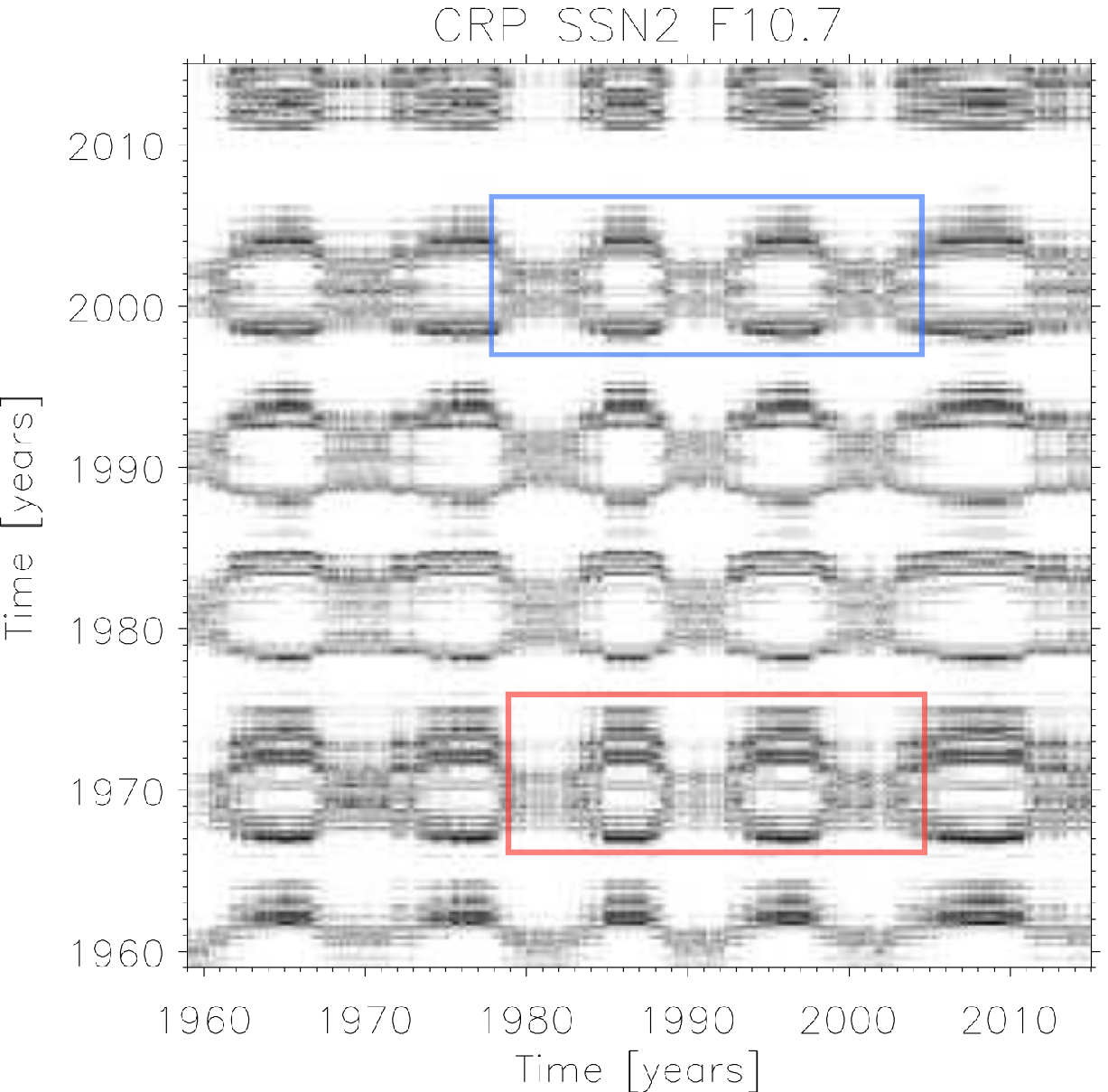}}
 \caption{JRPs of the SSN1 and F10.7 (upper panel), and SSN2 and F10.7 (lower panel). The boxes highlight different epochs in the JRPs where a clear difference between the two versions of SSN are observed (see text for more details).} 
 \label{CRP_SSN_F107}
 \end{figure}

With the aim of singling intrinsic variations of the system out of data artifacts, in Fig. \ref{CRP_SSN_F107} we also show the JRPs obtained from F10.7 and the previous SSN1 data series (upper panel), and from the F10.7 and the new SSN2 (lower panel). It is important to note that the two JRPs show some difference at specific epochs. More in particular, in the figure we highlight two specific regions (blue and red boxes respectively) where the JRP of SSN1 and F10.7 shows a smaller number of simultaneous recurrences between the two time series. In comparison, the JRP of SSN2 and F10.7 (see lower panel of the same figure) does show a more homogeneous distribution of simultaneous recurrence points in the same regions selected. This quantifies the effect of the recent recalibration of the SSN sequence as seen into the phase space.

\section{Discussions}
\label{disc}
The results derived from our RQA of the two solar cycle indices show that the determinism of the system represented by these two data series undergoes rapid fluctuations in time. This behaviour is consistent with previous findings by \citet{2002ESASP.506..197P}, who explained this fact in terms of variation of the intermittency in the ascending and descending phases of the solar activity. In our study we have also extended this analysis to the laminarity of the system. After filtering out the low-frequency solar cycle modulation (eleven-year period), both LAM and DET show a modulation. While there exist some degree of correlation between LAM and DET, we have identified periods at which these two RQA measures present a different behaviour. Interestingly, one of them corresponds with last solar minimum, which has been longer than the previous ones. More in detail, during this times, the increase of LAM (increase level of disorder) is not accompanied by a similar increase of DET, suggesting an overall increase of the laminarity of the system.\\
Besides, a randomization test of the data sequences has shown that most of the peaks of DET and LAM, including those appearing during the last minimum, are statistically significant, with a confidence level exceeding $95 \%$. These large fluctuations may suggest the presence of dynamical transitions \citet{marwan2013recurrence}. However, \citet{schinkel2008selection} noted that determinism does not relate exactly to the mathematical notion of the term, but rather underlines that deterministic processes have usually a larger number of diagonal lines in RPs, if compared to purely stochastic processes. This may also explain the periods at which the increase of DET and LAM is synchronous. In this regard, \citet{marwan2013recurrence} have also argued that, in some physical systems, the increase of the measure of DET together with that of LAM can be understood as a slowing down of the dynamics, typical of tipping points. These two facts together solve the apparent contradiction of the simultaneous increase of LAM and DET.\\ 
It is worth stressing that these results are obtained independently of the filtering technique (either FFT or EMD) adopted to remove the low-frequency modulation of the solar cycle indices.\\ 
However, the results of the RQA include much more information than mentioned. Indeed, one of the most clear indication emerging from them, is represented by the different dynamical behaviour of the two solar cycle indices investigated (SSN and F10.7). This is emerging not only from the temporal behaviour of the RQA measures of F10.7, whose peaks appear sharper than those obtained from SSN2, but also from the power spectra of DET and LAM fluctuations. In fact, while the power spectrum of the RQA measures of SSN2 presents power up to frequency of $1.7-2 \times 10^{-3}$ days$^{-1}$ (or equivalently periods in the range   $1-2$ years), the power spectrum of both LAM and DET of F10.7 appear limited to frequencies smaller than $\sim 5 \times 10^{-4}$  days$^{-1}$. These differences between the two indicators are not surprising. As already mentioned before, among the two, only F10.7 represents a physical quantity, SSN being the weighted count of the sunspots appearing on the solar disk over time. However, it is worth noting that the RQA provides a quantification of these differences, which in our opinion is helpful for uncovering the intrinsic meaning of the SSN. \\
In addition to this, in this work we also tested our results against the impact of the new SSN calibration. Although the above dynamical transitions occurring at the minima of the solar cycle are not sensitive to calibration issues, the JRP also reveals a significant number of discrepancies between the two data series which are, according to us, value-added results with respect to the main scope of this work, providing information that can be of more general interest to the community. Indeed, since our analysis preserves the non-linearities of the process, this comparison provides useful insight into the relationships between the SSN1 and SSN2, which are the most used solar index so far, and the one that will likely be the most commonly used in the future, respectively.\\
In more detail, although on average the new data series appears more deterministic, there exist specific times at which the SSN1 is characterized by less simultaneous recurrences with the F10.7 than the SSN2.
This witness the improvements made by the SSN2 over the former SSN1 \citep{2015arXiv151006928F}, although the residual differences between the F10.7 and SSN2 may offer good reasons for further working on the revision of available series. However, it is useful to remark once again that our analysis method is appropriate for the analysis of the properties of non-linear systems that might be particularly difficult to unveil with other techniques. For this reason the identification of differences between the SSN1, SSN2, and F10.7 in terms of recurrence states may provide useful complementary information with respect to other techniques. We also note that this topic deserves more attention and a more complete comparative analysis of other solar indices, which is beyond the scope here. This will be addressed in a future work. 

\section{Summary and Conclusions}
In this work we have shown the results of the application of the RQA on two indices of the solar cycle, namely the SSN and the F10.7. The RQA is nowadays a widely used technique to investigate non-linear dynamical systems and their transitions, yet not fully exploited to investigate the solar activity cycle. The RPs, as well as the RQA measures demonstrate the non-stationarity of the system governing the activity cycle itself, with a strong modulation of the RQA measures, and fluctuations that may be linked to dynamical phase transition. Besides, the RQA unveils significant differences in the dynamics of SSN and F10.7, which reflects their different physical nature. \\
Furthermore, our application of the RQA to both SSN1 and SSN2 provides a timely non-linear comparison between the newly recalibrated SSN data series and the previous one. Indeed, we found that SSN2, the new series, shows a larger degree of determinism with respect to the SSN1. 
Although this analysis was mainly performed to test our findings and the reliability of the transition markers of the dynamics, the results of the comparative study appear of more general validity and interest, providing a key-reading for reconsidering the past literature based upon the SSN1 data series, in light of its recalibration.

\begin{acknowledgements}
This study received funding from the European Unions Seventh Programme for Research, Technological
Development and Demonstration, under the Grant Agreements of the eHEROES
(n 284461, www.eheroes.eu), SOLARNET (n 312495, www.solarnet-east.eu), and SOLID
(n 313188, projects.pmodwrc.ch/solid/) projects. It was also supported by COST Action
ES1005 TOSCA (www.tosca-cost.eu), by the Istituto Nazionale di Astrofisica, and the MIUR-PRIN grant 2012P2HRCR on “The active Sun and its effects on Space and Earth climate". Sunspot data from the World Data Center SILSO, Royal Observatory of Belgium, Brussels.
\end{acknowledgements}

\end{document}